\documentclass[10pt,peerreview,draftcls]{IEEEtran}
\usepackage{times}		
\usepackage{citesort}
\usepackage{amssymb,amstext,amsmath}
\usepackage{bm}
\usepackage{graphicx}
\usepackage{url}
\usepackage{algorithm}
\usepackage[noend]{algpseudocode}
\usepackage{subcaption}
\usepackage{localmacros}
\usepackage{setspace}

\title{Sparse Estimation with the Swept Approximated Message-Passing Algorithm}

\author{Andre~Manoel,~%
	\thanks{A. Manoel is with the Institute of Physics, University of S{\~ a}o Paulo, %
			R. do Mat{\~a}o 187, 05508-090 S{\~ a}o Paulo, Brazil.}%
Florent~Krzakala,~%
	\thanks{F. Krzakala is with Universit\'e Pierre et Marie Curie and %
			\'Ecole Normale Sup\'erieure, 24 rue Lhomond, 75005 Paris, France.}%
Eric~W.~Tramel,~%
	\thanks{E. W. Tramel is with \'Ecole Normale Sup\'erieure, 24 rue Lhomond, 75005 Paris, France.}%
Lenka~Zdeborov\'a%
	\thanks{L. Zdeborov{\' a} is with Institut de Physique Th\'eorique, CEA Saclay, %
			and CNRS URA 2306, 91191 Gif-sur-Yvette, France.}
}

\def\(({\left(}
\def\)){\right)}
\def\[[{\left[}
\def\]]{\right]}

\begin{document}

\maketitle

\begin{abstract}
Approximate Message Passing (AMP) has been shown
to be a superior method for inference problems, such
as the recovery of signals from
sets of noisy, lower-dimensionality measurements, both
in terms of reconstruction accuracy and in computational
efficiency. However, AMP suffers from serious convergence
issues in contexts that do not exactly match its assumptions.
We propose a new approach to stabilizing AMP in these
contexts by applying AMP updates to individual coefficients 
rather than in parallel. Our results show that this change
to the AMP iteration can provide theoretically expected, but
hitherto unobtainable, performance for problems on which the
standard AMP iteration diverges. Additionally, we find that
the computational costs of this \emph{swept} coefficient update
scheme is not unduly burdensome, allowing it to be applied 
efficiently to signals of large dimensionality.
\end{abstract}


\section{Introduction}
\label{sec:intro}

Belief Propagation (BP) is a powerful iterative message passing
algorithm for graphical models \cite{Pea1988,MM2009,OS2001}. However,
it presents two main drawbacks when applied to highly connected
continuous variable problems: first, the need to work with continuous
probability distributions; and second, the necessity to iterate over
one such probability distribution for each pair of variables.

The first problem can be addressed by projecting the distributions onto a
finite number of moments \cite{sudderth2010nonparametric} and the second by
utilizing the Thouless-Andreson-Palmer (TAP) approach \cite{MM2009,OS2001}
where only single variable marginals are required. Approximate message
passing (AMP), first introduced in \cite{DMM2009}, is one
relaxation of BP that utilizes both of the aforementioned 
approximations in order to 
solve sparse estimation problems. In AMP's more general setting, as is
considered in Generalized AMP (GAMP) \cite{Ran2011}, the goal of the algorithm
is the
reconstruction of an $N$-dimensional sparse vector $\bf x$ given the
knowledge of an $M$-dimensional vector $\bf y$ obtained via a possibly
non-linear and/or probabilistic output function $h(z)$ performed on a
set of linear projections. Specifically,
\begin{equation}
y_{\mu} = h(z_{\mu}), \quad \text{where} \quad z_{\mu} = \sum_{i=1}^N \Phi_{\mu i}
x_i . \quad \footnotemark
\label{eq:forward_model}
\end{equation}
\footnotetext{In the present work, we use subscript notation to denote the individual coefficients of 
vectors, i.e. $y_{\mu}$ refers to the $\mu^{th}$ coefficient of ${\bf y}$ where 
$\mu \in \left\{1,2,\dots,M \right\}$, and the double-subscript notation to
refer to individual matrix elements in row-column order.}
For example, if $h(z)=z+\xi$
where $\xi$ is a zero-mean \iid Gaussian random variable, then $h(z)$ represents
an additive white Gaussian noise (AWGN) channel. With this output function, 
in the setting $M \ll N$, \eqref{eq:forward_model} is simply the application of
Compressed Sensing (CS) \cite{CR2005a} under noise. AMP is currently
acknowledged as one of the foremost algorithms for such problems in terms of both
its computational efficiency and in the number of measurements required for
exact reconstruction of {\bf x}. In fact, with
properly chosen measurement matrices \cite{KMS2012b,KMS2012,donoho2012information},
one can achieve information-theoretically optimal reconstruction performance for 
CS, a hitherto unachievable bound with standard convex optimization approaches.

Just as with any iterative algorithm, the convergence properties of
AMP are of chief analytical concern. Many rigorous results have been
obtained on the performance of AMP in the case of \iid and block \iid
matrices
\cite{BayatiMontanari10,donoho2012information}. Unfortunately,
while AMP performs well for zero-mean \iid projections, performance
tends to drastically decline if one moves away from these simple
scenarios. In fact, even for \iid matrices with a small positive mean,
the algorithm may violently diverge, leading to poor reconstruction
results \cite{CKZ2014}. This instability to slight variations from
these strict assumptions on the projections is a serious problem for
many practical applications of AMP.

The main theoretical reason for these convergence issues has been
identified in \cite{CKZ2014}. Namely, AMP's use of a parallel update, 
instead of a sequential one, on the BP variables at each iteration. 
Three strategies have been proposed in recent literature to avoid
this problem. 
First,
one can highly damp the AMP iterations, as in \cite{KMS2012b,VS2013}.
However, this often requires a
damping factor so large that the cost, in terms of the number of iterations
until convergence, is prohibitive. Additionally, it is not
entirely clear how to determine an optimal damping factor to ensure convergence
in general. 
Second, 
one can modify the problem
{\it a posteriori} in order to come back to a more favorable situation. For
instance, one might remove the mean of the matrix and of the
measurements \cite{CKZ2014}, or one might modify the algorithm according
to the theoretical
spectrum of the operator $\Phi$ \cite{RSF2014,SAMP2014}, if it is known. This
knowledge about the operator may be prohibitive and could
therefore present a strong limitation in practice. 
Third, one might take one step backward in approximation
from AMP to a BP-style iteration \cite{CKZ2014}. This amounts to a
huge cost in terms of both memory and computational efficiency 
as there are $O(N^2)$ variables to update with BP as opposed to the the 
$O(N)$ utilized in AMP.

In this contribution, we solve these problems by deriving a modified
and efficient AMP algorithm with greatly improved convergence
properties while preserving the $O(N)$ iteration and memory cost of
AMP. We accomplish this by a careful analysis of the relaxation leading from
BP to AMP where we preserve the sequential, or swept,
variable update pattern of BP in our AMP approach. This leads to a
slightly modified set of update rules for the AMP and GAMP algorithms without
affecting the fixed point in any way. The resulting algorithm, which we
denote as Swept AMP (\swamp), possesses
impressive empirical convergence properties. The derivation of \swamp~is
explained next in Sec.~\ref{sec:Derivingswamp}. We then report, in
Sec.~\ref{sec:Results}, numerical results for basic and 1-bit CS, 
as well as for group testing. In all of these cases, 
huge improvements over the state-of-the-art can be obtained while remaining
robust to projections with troublesome properties.


\section{From Belief-Propagation to \swamp~for Signal Recovery}
\label{sec:Derivingswamp}

\subsection{Signal Recovery as Statistical Estimation}
To describe AMP, we focus on the CS signal recovery problem with real
valued signals in terms of statistical inference. Given an unknown
signal $\vx \in \mathbb{R}^N$, a linear projection $\Phi \in
\mathbb{R}^{M \times N}$, and a set of observations $\vy \in
\mathbb{R}^M$ generated from $\vx$ and $\Phi$, we write the posterior
distribution for the unknown signal according to Bayes' rule,
\begin{equation}
    P (\vx | \Phi, \vy) \propto P (\vy | \Phi, \vx) \, P_0 (\vx),
    \label{eq:bg-post}
\end{equation}
where we write $\propto$ as we neglect the normalization constant. The
likelihood $P (\vy | \Phi, \vx)$ is determined according to the
constraints one wishes to enforce, which we consider to be of form
$\vy = h(\Phi \vx)$, with $h$ being, in general, any stochastic
function.  Here, we consider $h$ to be an
AWGN channel
\footnote{One can generalize $h$ to be a more complicated
output function. This generalization constitutes the change of AMP to
GAMP \cite{Ran2011}. For example, we examine the case of 1-bit CS in 
Sec. \ref{sec:onebitcs} where $h$ is a non-linear sign function.},
\begin{equation}
    y_{\mu} = h(\Phi_{\mu} \vx) = \Phi_{\mu} \vx + \mathcal{N} (0, \Delta),
\end{equation}
where $\Delta$ is the variance of the AWGN and $\Phi_{\mu}$ is the $\mu^{th}$ 
row-vector of $\Phi$. Hence, 
\begin{equation}
    P(\vy | \Phi, \vx) = \frac{1}{\sqrt{2 \pi \Delta}} \prod_{\mu =
    1}^M \exp \left[ - \frac{(y_\mu - \sum_i \Phi_{\mu i} x_i)^2}{2 \Delta}
    \right]. \label{eq:bg-llh}
\end{equation}

The prior $P_0 (\vx)$ is determined from the information we
have on the structure of $\vx$. For CS, we are concerned with
the recovery of {\it sparse} signals, i.e. ones with few non-zero
values. Unstructured sparse signals can be modeled well by an \iid~Bernoulli sparse
prior,
\begin{equation}
    P_0 (\vx) \propto \prod_{i = 1}^N P_0 (x_i), \quad \text{where} \quad P_0 (x_i) =
\rho \psi(x_i) + (1 - \rho) \delta(x_i),
\end{equation}
where $\psi (x_i)$ can be any distribution, e.g. $\psi (x_i) =
\mathcal{N} (x_i; \bar{x}, \sigma^2)$, and the degree of sparsity is controlled
by the value $\rho \in [0,1]$. Notice that, in this usual
setting, both distributions are factorized, that is, the likelihood is in $M$ terms
relative to the constraint over each $y_\mu$, and the prior is in $N$
terms relative to what is expected of each $x_i$. Factorized
distributions such as these are well represented by graphical models
\cite{WJ2008}, specifically, bipartite graphs in which the $M + N$ factors are
represented by one type of node and the $N$ variables $x_i$ by another.  
Once the posterior distribution is written
down, the estimate $\hat{\vx}$ may be assigned in different ways,
according to what loss function one wishes to minimize. In this work,
we are chiefly concerned with the minimum mean-squared error (MMSE) estimate, 
which can be shown to be the average of $x_i$ with respect to the posterior $P (\vx
| \Phi, \vy)$; if one were able to compute the posterior's marginals, the
MMSE estimate would read
\begin{equation}
\hat{x}_i^\text{MMSE} = \myint{x_i}{x_i P(x_i | \Phi, \vy)}, \quad \forall i.
\end{equation}
The strategy employed by AMP is to infer the
marginals of the posterior by using a relaxed version of the BP 
algorithm \cite{Pea1988,MM2009}, and thus to arrive at the MMSE estimate of
the unknown signal $\vx$.

\subsection{Relaxed Belief-Propagation}\label{sec:rBP}
BP implements a
message-passing scheme between nodes in a graphical model, ultimately
allowing one to compute approximations of the posterior marginals.
Messages $m_{i \to \mu}$ are sent from the variables nodes to the factor nodes
and subsequent messages $m_{\mu \to i}$ are sent from factor nodes back 
to variable nodes that corresponds to 
algorithm's current ``beliefs'' about the probabilistic distribution of the variables
$x_i$.
Since these distributions are continuous, the first relaxation step is to move to a
projected version of these distributions, as described in
\cite{Ran2011,KMS2012}.  Here, we shall follow the notation of reference
\cite{KMS2012} and use the following parametrization:
\begin{eqnarray} 
a_{i\to \mu } \definedas   \int {\rm d}x_i \, x_i \,  m_{i\to \mu}(x_i) \, &,&  \label{a_imu}
~~~v_{i\to \mu } \definedas \int {\rm d}x_i \, x^2_i \, m_{i\to \mu}(x_i) -
a^2_{i\to \mu } \, , \label{c_imu} \\
\quad m_{\mu \to i}(x_i) &\propto&e^{-\frac{x^2_i}{2}A_{\mu\to i} +  B_{\mu \to i} x_i}\, . \label{m_mui} 
\end{eqnarray} 
This leads (see \cite{KMS2012}) to the following closed recursion sometimes called relaxed
BP (r-BP):
%
\begin{align} 
&A_{\mu\to i} = \frac{\Phi^2_{\mu i}}{\Delta + \sum_{j\neq i} \Phi^2_{\mu j} v_{j\to \mu}}  \, , 
&&a_{i\to \mu}=\fa\left(\frac{1}{\sum_{\gamma\neq \mu}A_{\gamma \to
      i}},\frac{\sum_{\gamma\neq \mu}B_{\gamma \to i}}{
    \sum_{\gamma\neq \mu}A_{\gamma \to
      i}   }\right)\, , \\
&B_{\mu \to i} = \frac{\Phi_{\mu i}(y_\mu - \sum_{j\neq i} \Phi_{\mu
    j}a_{j\to \mu})}{\Delta + \sum_{j\neq i} F^2_{\mu j} v_{j\to
    \mu}} \, , 
&&v_{i\to \mu}=\fc\left(\frac{1}{\sum_{\gamma\neq \mu}A_{\gamma \to
      i}},\frac{\sum_{\gamma\neq \mu}B_{\gamma \to i}}{
    \sum_{\gamma\neq \mu}A_{\gamma \to
      i}   }\right)  \, ,\label{eqq2}
\end{align} 
where the functions $f$ are defined by  the following prior-dependent
integrals
\begin{eqnarray} 
    \fa(\Sigma^2,R) &\definedas& \int {\rm d}x\,  x \,   P_0 (x)\frac{1}{\sqrt{2\pi} \Sigma} e^{-\frac{(x-R)^2}{2\Sigma^2}}\, , \label{f_a_gen}\\
    \fc(\Sigma^2,R) &\definedas& \int {\rm d}x\,  x^2 \,   P_0 (x)\frac{1}{\sqrt{2\pi} \Sigma} e^{-\frac{(x-R)^2}{2\Sigma^2}} -
    \fa^2(\Sigma^2,R) = \Sigma^2  \frac{{\rm d}\fa}{{\rm d} R}
      (\Sigma^2,R) \, .\label{f_c_gen}
\end{eqnarray} 
After convergence, the single
point marginals are given by
\begin{eqnarray} 
a_{i}&=&\fa\left(\frac{1}{\sum_{\gamma}A_{\gamma \to
      i}},\frac{\sum_{\gamma}B_{\gamma \to i}}{ \sum_{\gamma}A_{\gamma
      \to
      i}   }\right) \, , \label{eqq3}~~~~
v_{i}=\fc\left(\frac{1}{\sum_{\gamma}A_{\gamma \to
      i}},\frac{\sum_{\gamma}B_{\gamma \to i}}{ \sum_{\gamma}A_{\gamma
      \to i} }\right)\, .  \label{eqq4}
\end{eqnarray} 
We intentionally write r-BP without specifying time indices since the
updates can be performed in one of two ways. The first approach is to
update in parallel, where {\it all} variables are updated at time $t$
given the state at time $t-1$. The second is the random sequential
update where one picks a {\it single} index $i$ and updates all
messages corresponding to it. A time-step is completed once all
indices have been visited and updated once. As shown in
\cite{CKZ2014}, the sequential, or swept, iteration is much more
stable for r-BP. We now turn our attention to AMP and to our proposed
modification.
\subsection{Swept Approximate Message Passing} \label{sec:amp}
In the message-passing described in the previous section, 
$2 (M\times N)$ messages are
sent, one between each variable component and each measurement at each
iteration. This creates a very large computational and memory burden
for applications with large $N$, $M$.  It is possible to rewrite the
BP equations in terms of only $N+M$ messages by making the assumption
that $\Phi$ is dense and that its elements are of magnitude
$O(1/\sqrt{N})$.  In statistical physics, this assumption leads to the
TAP equations \cite{TAP1977} used in the study of spin glasses.  For
graphical models, such strategies have been discussed in
\cite{OS2001}. The use of TAP with r-BP provides the standard AMP
iteration, as we now show. First we define
%
\begin{align}
  &\omega_\mu \definedas \sum_i \Phi_{\mu i} a_{i \to\mu}, \quad \quad \quad \quad &&V_{\mu} \definedas \sum_i \Phi_{\mu i}^2  v_{i \to\mu}, \label{OG}\\
  &\Sigma^2_i \definedas \frac{1}{\sum_{\mu}A_{\mu\to i}}, \quad \quad \quad \quad &&R_i     \definedas \frac{\sum_{\mu}B_{\mu\to i}} {\sum_{\mu}A_{\mu\to i}}. \label{UV}
\end{align}
Next we expand around the marginals and disregard any $O(1)$ terms 
(see \cite{CKZ2014} for details)
to find:
\begin{eqnarray}
V_{\mu} &\approx& \sum_i \Phi_{\mu i}^2 v_i, \, ,\nonumber \\
\Sigma^2_i\!\!&=&\!\! \left[ \sum_\mu \frac{\Phi^2_{\mu i}}{\Delta_\mu + V_\mu - \Phi^2_{\mu
    i} v_{i\to \mu}} \right]^{-1} \approx \left[ \sum_\mu \frac{\Phi^2_{\mu i}}{\Delta_\mu +
  V_\mu} \right]^{-1} \, , \nonumber \\
R_i\!\!&=&\!\!  \left[  \sum_\mu
\frac{\Phi_{\mu i}(y_\mu - \omega_\mu+ \Phi_{\mu i}a_{i\to \mu})}{\Delta_\mu + V_\mu - \Phi^2_{\mu
    i} v_{i\to \mu}}\right] \left[ \sum_\mu \frac{\Phi^2_{\mu i}}{\Delta_\mu + V_\mu - \Phi^2_{\mu
    i} v_{i\to \mu}} \right]^{-1} ,\nonumber \\
&\approx&  a_i + \frac{
\sum_\mu \Phi_{\mu i}
\frac{(y_\mu - \omega_\mu)}{\Delta_\mu +V_\mu}}{  \sum_\mu \Phi_{\mu i}^2
\frac{1}{\Delta_\mu + V_\mu}   }
\, .
\end{eqnarray}
Now let us investigate the expansion of the factor $\omega_{\mu}$ as we
include the time, or iteration, indices $t$. First one has
\begin{eqnarray}
  a_{i\to \mu}^{t+1} &=& \fa\left(\frac{1}{\sum_{\nu}A^t_{\nu\to i}
      -A^t_{\mu\to i}},\frac{ \sum_{\nu}B^t_{\nu\to i} -B^t_{\mu\to
        i}}{ \sum_{\nu}A^t_{\nu\to i} -A^t_{\mu\to i} }\right)=
a_i^{t+1} - B^{t}_{\mu\to i}  (\Sigma_i^2)^t \frac{\partial
  \fa}{\partial R} \left((\Sigma_i^2)^t,R_i^t\right), \nonumber \\
&=& a_i^{t+1} - B^{t}_{\mu\to i}   v_i^{t+1},
\end{eqnarray}
making the expansion for $\omega_{\mu}$ 
\begin{eqnarray}
\omega_\mu^{t+1}
= \sum_i \Phi_{\mu i} a_i^{t+1} -\frac{
  (y_\mu-\omega_\mu^t)}{\Delta_\mu +V_\mu^t} \sum_i \Phi_{\mu i}^2
v_i^{t+1}
= \sum_i \Phi_{\mu i} a_i^{t+1} -\frac{
  (y_\mu-\omega_\mu^t)}{\Delta_\mu +V_\mu^t} V_{\mu}^{t+1}
 \, , \label {eq:w}
\end{eqnarray}
which allows us to close the equations on the set of $a,v,R,\Sigma,V$ and
$\omega$. Iterating all relations in parallel (i.e. updating all
$R,\Sigma$'s, then $a,v$'s and then the $\omega,V$'s) provides the
AMP iteration. 

The implementation of the sequential update is not a straightforward task as
many otherwise intuitive attempts lead to non-convergent algorithms.
The key observation in the derivation of SwAMP is that \eqref{eq:w} mixes different time
indices: while the ``$a$'' and ``$V$'' are the ``new ones'', the
expression in the fraction is the ``old'' one, i.e. the one {\it
  before} the last iteration.  The implication of this is that while $
\sum_i \Phi_{\mu i} a_i$ and $V_\mu$ should be recalculated as the
updates sweep over $i$ at a single time-step, the term
$(y_\mu-\omega_\mu)/(\Delta_\mu +V_\mu)$
(which we denote as $g_{\mu}$ later on) should not. A corresponding
bookkeeping then leads to the \swamp~algorithm for the evolution of
$\omega_\mu$, $\Sigma^2_i$, $ V_{\mu}$ and $R_i$ described
in Alg.~\ref{alg:amp}. At this point, the difference between AMP and
\swamp~appears minimal, but, as we shall see, the differences in
convergence properties turn out to be spectacular.

\begin{algorithm}[h]
{\fontsize{8}{8}\selectfont
\begin{algorithmic}[1]
    \Procedure{swamp}{$y$, $\Phi$, $\{\Delta, \theta_\text{prior}, t_\text{max}, \varepsilon\}$}
    \State $t \gets 0$
    \State initialize $\left\{ \va^{(0)},\vv^{(0)} \right\}$, 
    $\{ \vw^{(0;\,N+1)},\vV^{(0;\,N+1)}\}$
    \While{$t < t_\text{max}$ \AND $|\!| \va^{(t + 1)} - \va^{(t)} |\!| < \varepsilon$}
        \For{$\mu = 1, M$}
            \State $g_\mu^{(t)}                       \gets \frac{y_\mu - \omega_\mu^{(t;\,N+1)}}{\Delta + V_\mu^{(t;\,N+1)}}$
            \State $V_\mu^{(t+1;\,1)}                   \gets \sum_i \Phi^2_{\mu i} v_i^{(t)}$
            \State $\omega_\mu^{(t+1;\,1)}              \gets \sum_i \Phi_{\mu i} a_i^{(t)} - V_\mu^{(t+1;\,1)} g_\mu^{(t)}$
        \EndFor
        \State $\mathbf{S} \gets \text{Permute}([1,2,\dots,N])$
        \For{$k = 1, N$}
            \State $i \gets S_k$
            \State ${\Sigma_i^2}^{(t+1)}              \gets \left[\sum_{\mu} \frac{\Phi^2_{\mu i}}{\Delta + V_{\mu}^{(t+1;\,k)}} \right]^{-1}$
            \State $R_i^{(t+1)}                       \gets a_i^{(t)} + {\Sigma_i^2}^{(t+1)} \sum_{\mu} \Phi_{\mu i} \frac{y_{\mu} - \omega_\mu^{(t+1;\,k)}}{\Delta + V_{\mu}^{(t+1;\,k)}}$
            \State $a_i^{(t+1)}                       \gets \fa(R_i^{(t+1)},{\Sigma_i^2}^{(t+1)}; \theta_\text{prior})$
            \State $v_i^{(t+1)}                       \gets \fc(R_i^{(t+1)},{\Sigma_i^2}^{(t+1)}; \theta_\text{prior})$
            \For{$\mu = 1, m$}
                \State $V_\mu^{(t+1;\,k+1)}              \gets V_\mu^{(t+1;\,k)} + \Phi_{\mu i}^2 \, (v_i^{(t+1)} - v_i^{(t)})$
                \State $\omega_\mu^{(t+1;\,k+1)}         \gets \omega_\mu^{(t+1;\,k)} + \Phi_{\mu i} \, (a_i^{(t+1)} - a_i^{(t)}) - g_\mu^{(t)} (V_\mu^{(t+1;\,k+1)} - V_\mu^{(t+1;\,k)})$
            \EndFor
        \EndFor
        \State $t \gets t + 1$
    \EndWhile
    \EndProcedure
    \State \textbf{return} $\left\{ \va^{(t+1)},\vv^{(t+1)} \right\}$
\end{algorithmic}
}
\caption{Swept AMP}
\label{alg:amp}
\end{algorithm}

Finally, we note that this procedure can also be generalized, a la
GAMP, 
for output channels other than the AWGN.
The required change is minimal \cite{Ran2011}: one should replace the
term $(y_\mu-\omega_\mu)/(\Delta_\mu +V_\mu)$ in the $R_i$ and
$\omega_\mu$ updates with $g_\text{out} \left( \omega_\mu, V_\mu
\right)$, a generic function which depends on the
channel. Specifically, $g_\text{out} (\omega, V) = \int dz \, P(y | z)
\, e^{- \frac{(z - \omega)^2}{2 V}} \left( \frac{z - \omega}{V}
\right)$.  Additionally, the $\frac{1}{\Delta_\mu + V_\mu}$ term in
the $\Sigma_i^2$ update should be replaced by $-\frac{\partial
  g_\text{out}}{\partial \omega}$. Notice that all AWGN specific
terms are recovered for $P (y | z) \propto e^{- \frac{(y - z)^2}{2
    \Delta}}$.


\section{Numerical Results}
\label{sec:Results}
Here, we present a range of numerical results demonstrating the effectiveness
of the \swamp~algorithm for problems on which both standard AMP and 
$\ell_1$ minimization via convex optimization fail to provide desirable 
reconstruction performance. All experiments were conducted on a computer
with an i7-3930K processor and run via Matlab. We have provided 
demonstrations of the \swamp~code on-line\footnote{\url{https://github.com/eric-tramel/SwAMP-Demo}}. 
For calculating $\ell_1$ recoveries, we utilize an implementation of the
SPGL1 \cite{BF2008} algorithm\footnote{\url{http://www.cs.ubc.ca/~mpf/spgl1/}}.


\subsection{Compressed Sensing with Troublesome Projections}
\begin{figure}[ht]
    \begin{subfigure}[b]{0.49\figurewidth}
        \includegraphics[width=\textwidth]{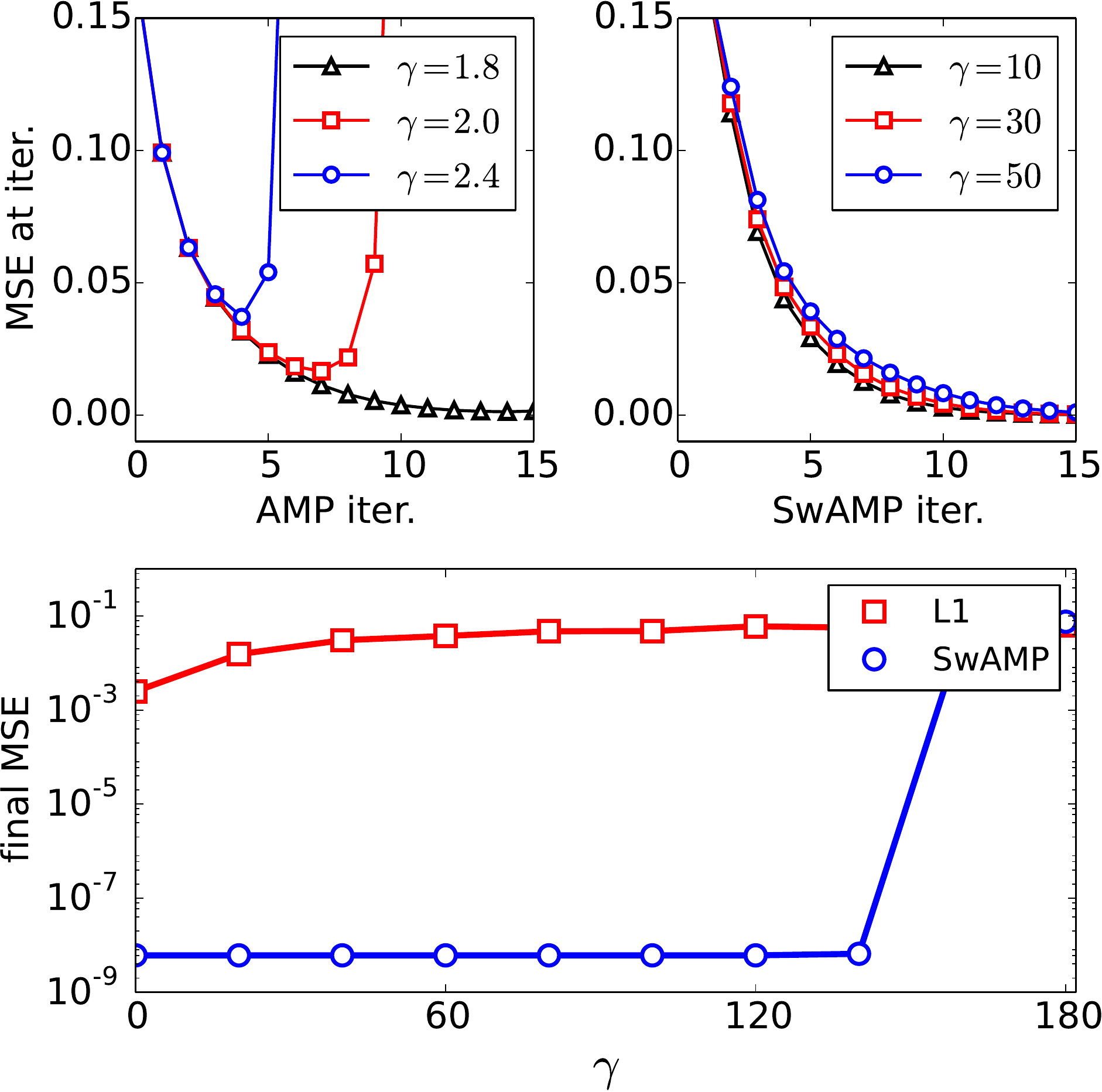}
        \caption{}
        \label{fig:diag_nzm}
    \end{subfigure}
    \begin{subfigure}[b]{0.49\figurewidth}
        \includegraphics[width=\textwidth]{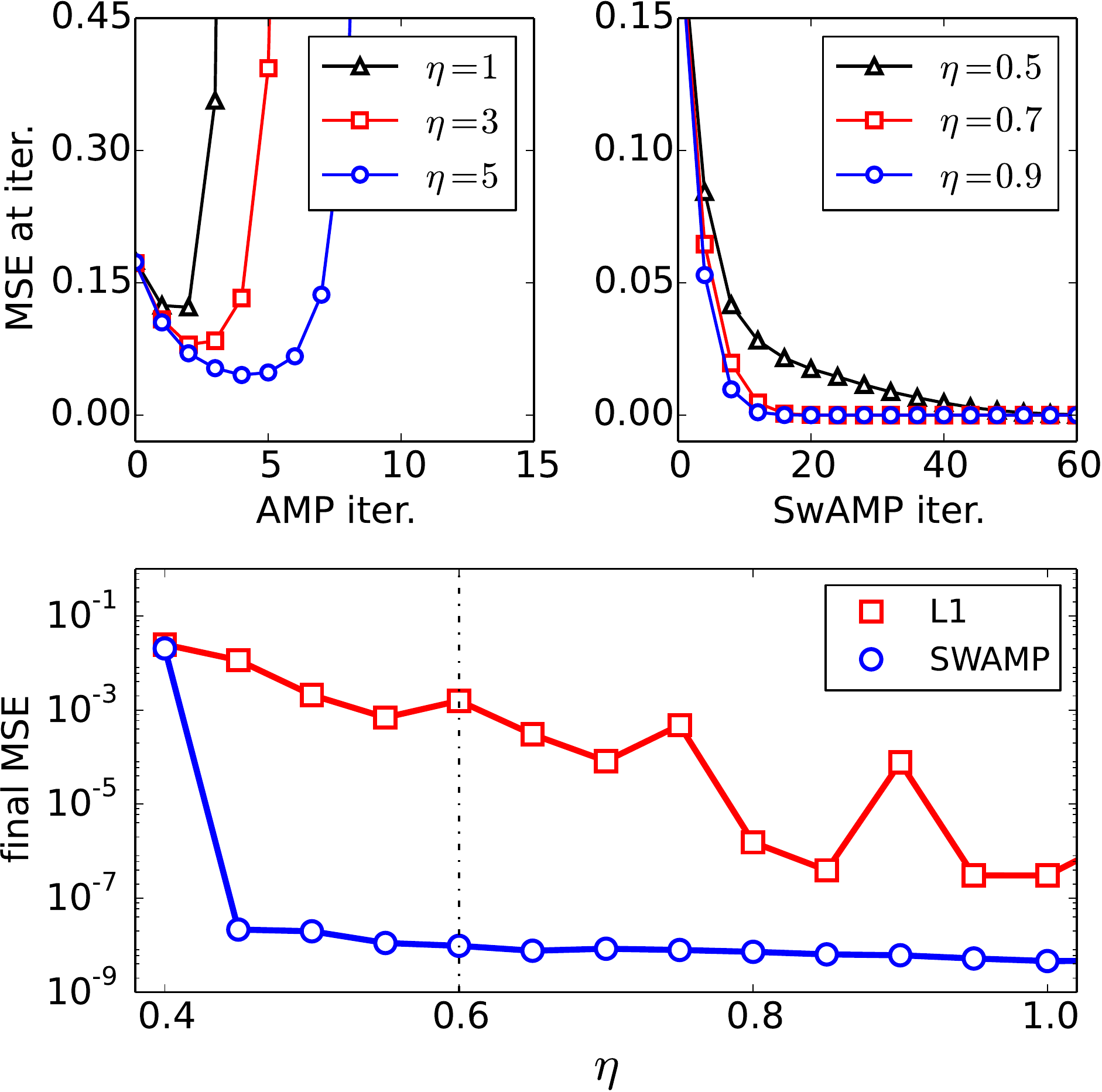}
        \caption{}
        \label{fig:diag_lr}
    \end{subfigure}
    \caption{AMP, \swamp, and $\ell_1$ solvers compared for
      CS signal reconstruction for sensing
      matrices with positive mean (left, a) and of low-rank
      (right, b) on sparse signals of size $N = 10^4$
      and sparsity $\rho=0.2$ with
      noise variance $\Delta = 10^{-8}$. The projections
      for (a) have been
      created following \eqref{eq:nonzero} using $M=\alpha N$
      measurements with $\alpha = 0.5$. The projectors for (b)
      have been created according to \eqref{eq:lowrank} and are low-rank
      for $\eta < \alpha=0.6$. 
      Finally, a comparison between reconstruction error obtained by \swamp~and
      $\ell_1$-minimization is given at the bottom of both (a) and (b) for the same experimental settings.}
\end{figure}

As discussed earlier, using projections of non-zero mean
to sample $\vx$ is one of the simplest cases for which AMP can fail to
converge. However, by using the proposed \swamp~approach, accurate estimates
of $\vx$ can be obtained even when the mean of the projections is 
non-negligible. While it may be possible to use
mean subtraction,
our proposed approach does not require such preprocessing. 
Additionally, as we will show later, not
all problems are amenable to such mean subtraction. 
To evaluate the effectiveness of \swamp~as compared to 
the standard parallel-update AMP iteration, we draw \iid~projections according
to
\begin{equation}
    \Phi_{\mu i} \sim \mathcal{N} \left( \frac{\gamma}{N}, \frac{1}{N}
    \right),
\label{eq:nonzero}
\end{equation}
where the magnitude of the projector mean is controlled by the term $\gamma$. 
For a given signal $\vx$ and noise variance $\Delta$, as $\gamma$ increases
from 0, we expect to see AMP failing to converge. This behavior can be observed in
the numerical experiments presented in Fig.~\ref{fig:diag_nzm}. Here, we 
observe that \swamp~is robust to values of $\gamma$ over an order of magnitude
larger than the standard AMP, converging to a low-MSE solution even for 
$\gamma \approx 140$ while AMP fails already at $\gamma = 2$. Additionally, for the tested parameters, $\ell_1$ minimization fails to provide a meaningful
reconstruction for any value of $\gamma$.

We also considered an even more troublesome case for projections, namely,
a set of projections which are strongly correlated. For these tests, we
draw 
\begin{equation}
    \Phi = \frac{1}{N} P Q, \quad \text{where}\quad P_{\mu k}, Q_{k i} 
    \sim \mathcal{N}(0,1)
\label{eq:lowrank}
\end{equation}
with $P \in \mathbb{R}^{M \times R}$, $Q \in \mathbb{R}^{R \times N}$ and $R
\triangleq \eta N$. That is, $\Phi$ is \emph{low-rank} for $\eta < \alpha$, 
where $\alpha = \frac{M}{N}$. In our experiments, we use $\eta$
to denote the level of independence of the rows of $\Phi$, with lower
values of $\eta$ representing a more difficult problem.
We observe that the elements of $\Phi$ are neither normal nor \iid~for 
these experiments. In Fig.~\ref{fig:diag_lr} we see that \swamp~is robust
to even these extremely troublesome projections while 
AMP fails to converge and $\ell_1$ minimization does not provide the
same level of accuracy as \swamp. These two experiments demonstrate how
the proposed \swamp~iteration allows for AMP-like performance while remaining
robust to conditions outside of the TAP assumptions about the projector.

\subsection{Group Testing}
Group testing, also known as {\it pooling} in molecular biology, 
is an approach to designing experiments so as 
to reduce the number of tests required to identify rare 
events or faulty items.
In the most naive approach to this problem, 
the number of tests is equal to the number of items, 
as each item is tested individually. However, since only
a small fraction of the items may be faulty, the number of tests can
be significantly reduced via pooling, i.e. 
testing many items simultaneously and allowing items to 
be included within multiple different tests. 
The nature of this linear combination of tests allows for a 
CS-type approach to faulty item detection, but
with a few important caveats. 
First, the operator is extremely sparse since the number of pools, 
and the number of items in them, may be limited due to physical
testing constraints.
Second, the elements of this operator are commonly $0/1$.
Group testing is therefore a very challenging application for AMP
since the properties of the group testing operator do not match 
AMP's assumptions.

\begin{figure}[ht]
    \begin{subfigure}[b]{0.4\figurewidth}
      \centering
      \includegraphics[width=\textwidth]{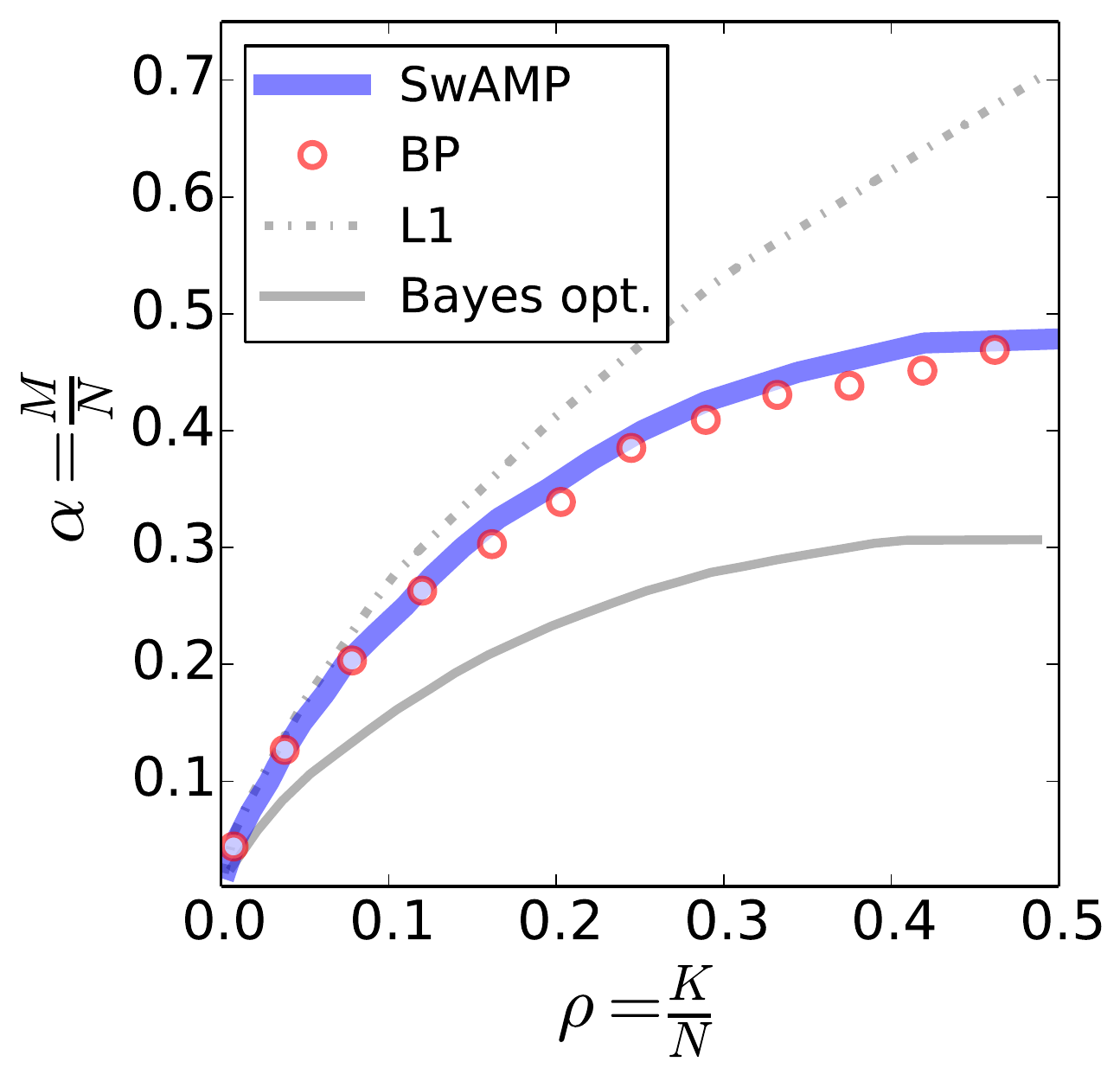}
      \caption{}
      \label{fig:diag_pool}
    \end{subfigure} \hspace{2em}
    \begin{subfigure}[b]{0.5\figurewidth}
      \centering
      \includegraphics[width=\textwidth]{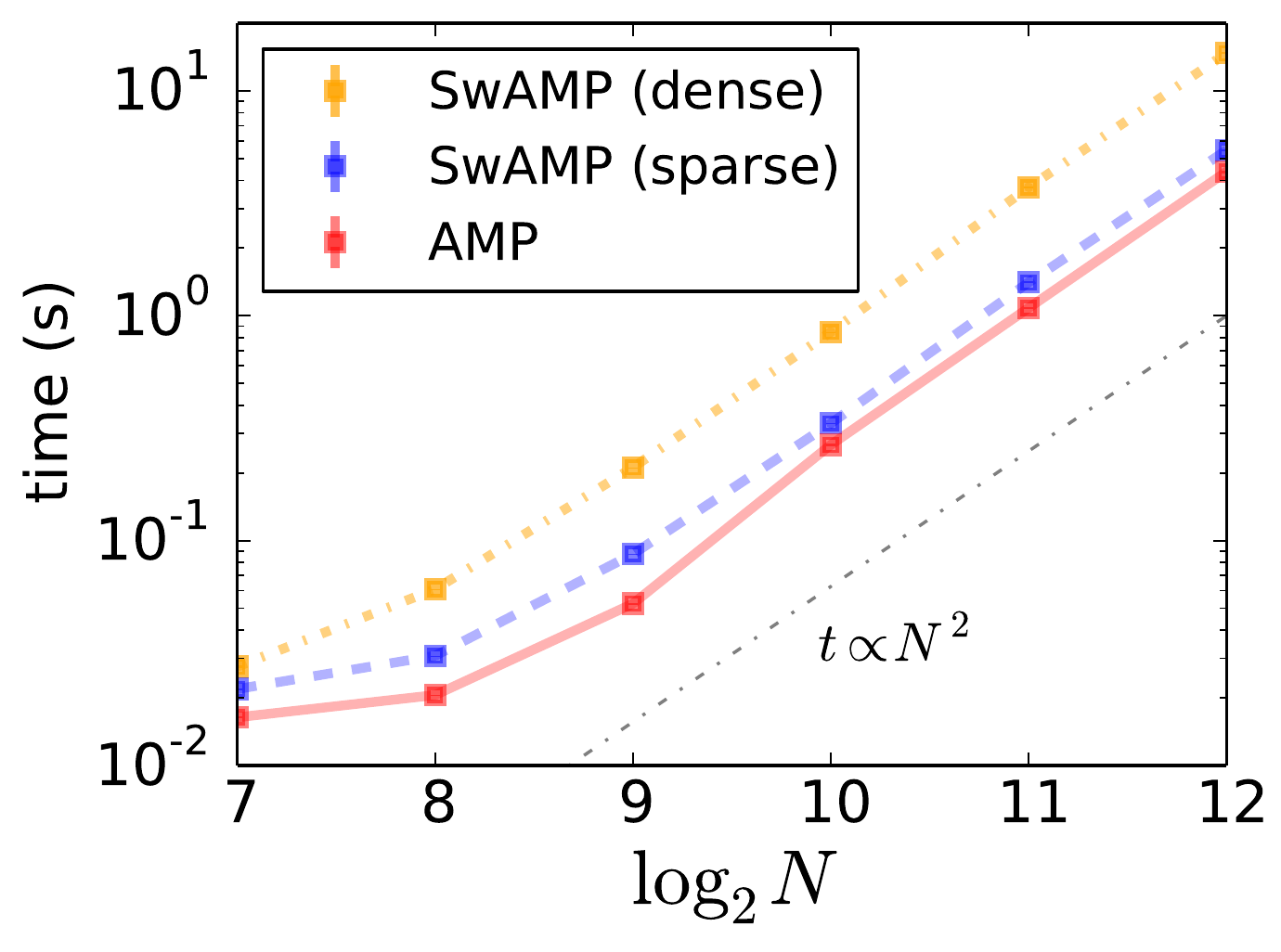}
      \caption{}
      \label{fig:benchmark}      
    \end{subfigure}
    \caption{%
    (a) 
    Group testing phase transition diagram between successful and unsuccessful signal
    recovery over $M$, the number of pools, and $K$, the number of non-zero
    signal elements. Successful recovery means the correct identification of all
    signal elements. The top-left of the diagram represents the easiest problems
    while the bottom-right the most difficult. The transition lines are drawn along the
    contour of 50\% of recoveries succeeding for many trials.
    (b) Execution times for both \swamp~and AMP using a sparse matrix
    with 25\% of its elements having non-zero value. 
    The reported times are measured for 500 iterations of the
    algorithms for each value of $N$ for the parameters $\rho = 0.25$ and
    $\alpha = 0.75$.}
\end{figure}

In one recent work \cite{ZKM2013}, the authors use both BP
and AMP for group testing and found that while basic AMP would not
converge, very good results---optimal ones, in fact---could
be obtained by using a BP approach. This came at
a large computational cost, however. Here, we have repeated the experiment of
\cite{ZKM2013} using the \swamp~approach instead of AMP and BP. In
fact, for \swamp, a sparse operator is a very advantageous situation
in terms of computational efficiency. Since the projector is extremely sparse
by construction, we may explicitly
ignore operations involving null elements, thus considerably improving
the algorithm's speed, as seen in Fig.~\ref{fig:benchmark}. Here, we also
see that \swamp's computational complexity is on the order of $O(N^2)$, as
is AMP's. 
Group testing experiments are shown in Fig.~\ref{fig:diag_pool} 
where we use random $0/1$ projections,
under the constraint that each projection should sum to $7$,
to sample sparse $0/1$ signals with $K \ll N$ non-zero elements,
where $N$ is the signal dimensionality. 
While AMP diverges when attempting to recover these signals, 
\swamp~converges to
the correct solution in few iterations. Additionally, 
\swamp~very closely matches the BP transition, thus providing 
recovery performance better than convex optimization, just
as BP does, but with much less computational complexity.

\subsection{1-bit Compressed Sensing}
\label{sec:onebitcs}

One of the confounding factors regarding the practical implementation
of CS in hardware devices is the treatment of measurement quantization.
The original CS analysis provides recovery bounds based
upon the assumption of real-valued measurements. However, in practice,
hardware devices cannot capture such values with infinite precision, and
so some kind of quantization on the measurements must be implemented.
Specifically, if $Q(\cdot)$ is
a uniform scalar quantizer, then $\vy = Q(\Phi \vx, B),$
where $B$ is the number of bits used to represent the measurement.
If signal recoverability is
significantly impacted by small $B$, then the dimensionality reduction
provided by CS may be lost by the requirement for many bits to encode
each measurement. 

Thankfully, recent works have shown CS recovery to be robust
to quantization and the non-linear error it introduces.
In fact, CS has been shown \cite{BB2008,JLB2011} to be robust even in
the extreme case $B=1$ known as 1-bit CS. 
In this case, the quantized measurements are given by
\begin{equation}
	\vy = \text{sign}(\Phi \vx).
    \label{eq:1bitcs}
\end{equation}
The non-linearity and severity of 1-bit CS requires special 
treatment from the CS recovery procedure. In \cite{BB2008}, 
a renormalized fixed-point continuation (RFPC) algorithm was proposed.
Later, \cite{JLB2011} analyzed the sensitivity of 1-bit CS to sign
flips and proposed a noise-robust recovery algorithm, binary
iterative hard thresholding (BIHT). 

\begin{figure}[ht]
    \centering
    \begin{subfigure}[b]{0.4\figurewidth}
        \includegraphics[width=\textwidth]{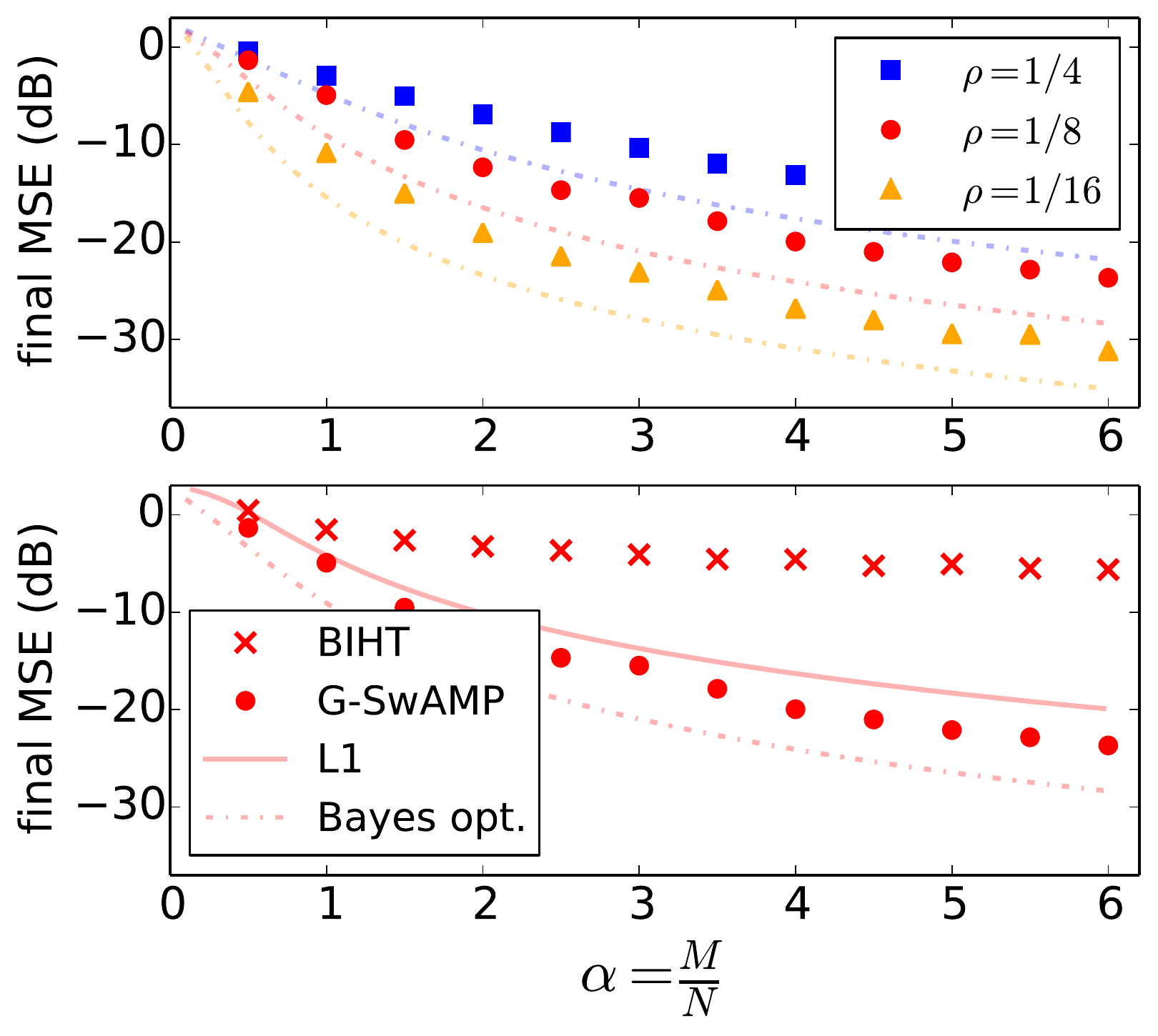}
        \caption{}
        \label{fig:1bit_all}
    \end{subfigure} \hspace{2em}
    \begin{subfigure}[b]{0.5\figurewidth}
        \includegraphics[width=\textwidth]{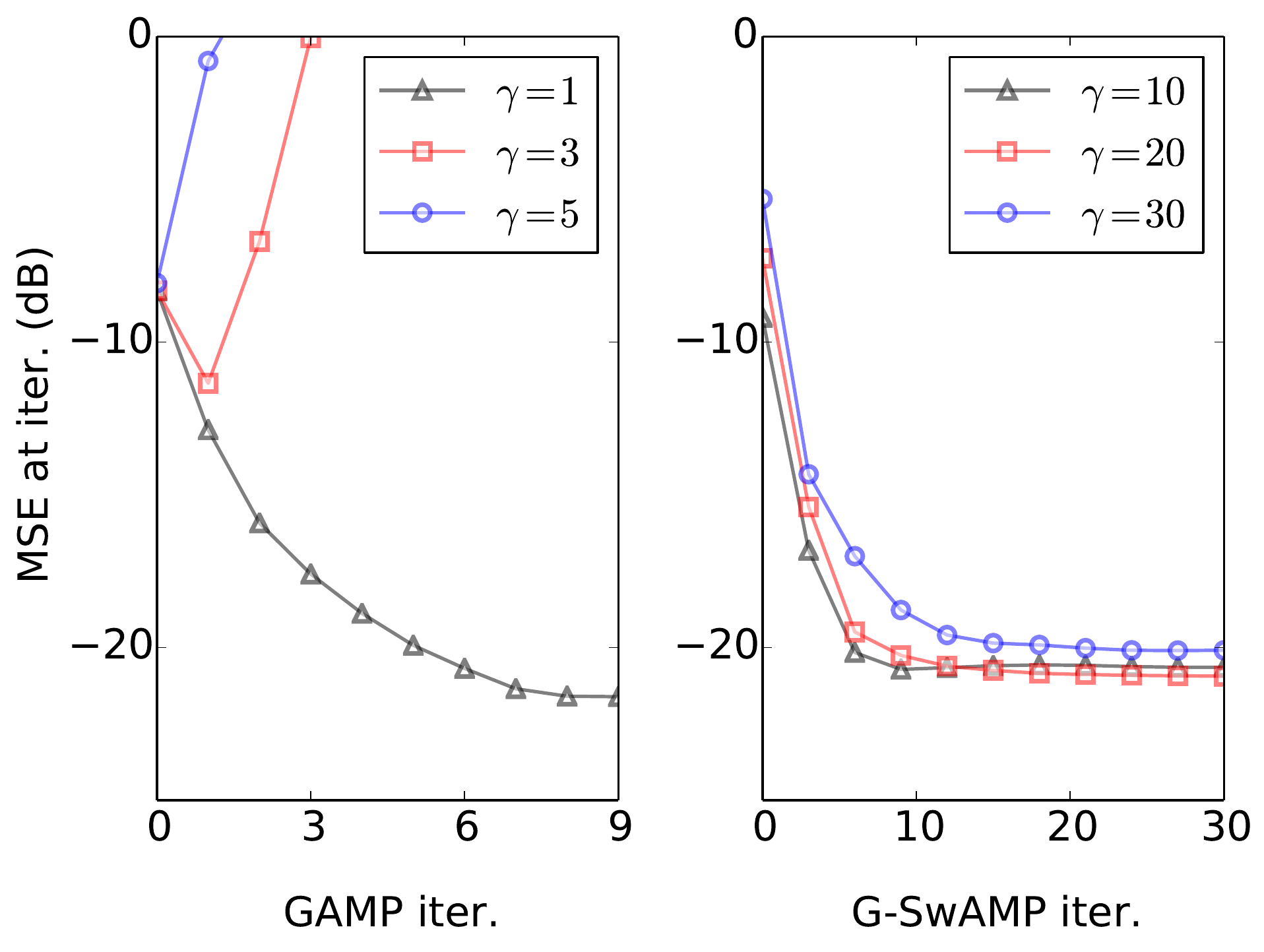}
        \caption{}
        \label{fig:1bit_vsgamp}
    \end{subfigure}
    \caption{Results for 1-bit CS. (a) Top: Comparison between the
      Bayes optimal MSE for zero-mean projectors
      \cite{XKZ2014} (dashed lines) and that obtained by \swamp~for
      projectors with $\gamma = 20$ (markers) for three
      different levels of signal sparsity. The reported empirical
      results were obtained by averaging over 200 instances of size $N
      = 512$. Bottom: Comparison of \swamp~and BIHT for $\rho = 1/8$
      for experiment conditions identical to the figure above; theoretical
      results for zero-mean projectors are also presented for completeness,
      including theoretical $\ell_1$ performance \cite{XK2013}. (b)
      Single instance comparison between GAMP and G-\swamp~ for 1-bit
      CS with $N = 2048$, $\rho = 1/8$, and $\alpha = 3$.}
\end{figure}

Recognizing the capability of GAMP to handle non-linear output
channels, \cite{KGR2012} proposed the use of GAMP for signal recovery
from quantized CS measurements. 
Further analysis of message-passing approaches to the
1-bit CS problem from the perspective of statistical mechanics was 
given in \cite{XK2013} where a modified fixed-point iteration was derived 
via the cavity method which provided both improved recovery accuracy and reconstruction time as compared to the RFPC. 
Additionally, the authors used replica analysis to 
estimate the optimal MSE performance of $\ell_1$-minimization based 
1-bit CS reconstruction. Finally, this analysis is extended in \cite{XKZ2014}
to include the theoretical Bayesian optimal performance, which we will use
as a baseline of comparison in Fig.~\ref{fig:1bit_all}.

Both methods \cite{KGR2012} and \cite{XK2013} show the
effectiveness of algorithms grounded in statistical mechanics for
quantized CS reconstruction. However, both assume an amenable set of
projections. Even projections possessing small mean can cause large
degradations in performance.
While mean removal is occasionally effective in the usual CS setting,
it cannot be used for 1-bit CS due to the nature of the sign operation
in \eqref{eq:1bitcs}. An algorithm that can handle troublesome
projectors can therefore be of great use.  In Sec. \ref{sec:amp}, we
show how the \swamp~can be modified to the general-channel setting, as
was done in GAMP. This generalization allows for 1-bit CS recovery
with \swamp~under much more relaxed requirements for $\Phi$.

In Fig.~\ref{fig:1bit_all},
we see Generalized \swamp~(G-\swamp) results for $\Phi_{\mu i} \sim \mathcal{N}
\left( \frac{20}{N}, \frac{1}{N} \right)$. 
We observe that G-\swamp~performs admirably even for this non-neglible 
mean on the projectors. In terms of recovery performance, it does not quite
meet the theoretical Bayes optimal performance \cite{XKZ2014}, however, this
is expected as the Bayes optimal performance is calculated for $\gamma = 0$.
Additionally, we see that even for this non-zero mean, G-\swamp~outperforms both
the BIHT's empirical performance for the same mean,
as well as the best-case theoretical $\ell_1$ performance for zero mean \cite{XK2013}. 
Finally, in Fig.~\ref{fig:1bit_vsgamp}, we see that
GAMP fails to provide any meaningful signal recovery for $\gamma$ small, while
G-\swamp~continues to converge to low-MSE even for large values of $\gamma$. 

\section{Conclusion}
\label{sec:conclusion}

While the AMP algorithm has been shown to be a very desirable
approach for signal recovery and statistical inference problems
in terms of both computational efficiency and accuracy, it is also
very sensitive to problems which deviate from its fundamental assumptions.
In this work, we propose the \swamp~algorithm which matches AMP's accuracy
while remaining robust to such variations, all without unduly increasing
computation or memory requirements. We also demonstrate how \swamp~can be
used to solve practical problems for which AMP and GAMP cannot be
applied, namely, group testing and 1-bit CS with troublesome projections. In
all cases, \swamp~provides superior accuracy as compared to $\ell_1$-
minimization, as well as convergence properties superior to AMP and GAMP, and all
with less computational and memory burden than BP or r-BP.

Exact analysis of the asymptotic state evolution of \swamp, as well as a
thorough analytical proof of its convergence, 
remains a challenging open problem for future work.

\section{Acknowledgments}
This work has been supported in part by the ERC under the European
Union’s 7th Framework Programme Grant Agreement 307087-SPARCS, by the
Grant DySpaN of ‘‘Triangle de la Physique,’’ and by FAPESP under grant
13/01213-8.

\newpage
\bibliographystyle{IEEEtran}
\bibliography{references}

\begin{thebibliography}{10}
\providecommand{\url}[1]{#1}
\csname url@samestyle\endcsname
\providecommand{\newblock}{\relax}
\providecommand{\bibinfo}[2]{#2}
\providecommand{\BIBentrySTDinterwordspacing}{\spaceskip=0pt\relax}
\providecommand{\BIBentryALTinterwordstretchfactor}{4}
\providecommand{\BIBentryALTinterwordspacing}{\spaceskip=\fontdimen2\font plus
\BIBentryALTinterwordstretchfactor\fontdimen3\font minus
  \fontdimen4\font\relax}
\providecommand{\BIBforeignlanguage}[2]{{%
\expandafter\ifx\csname l@#1\endcsname\relax
\typeout{** WARNING: IEEEtran.bst: No hyphenation pattern has been}%
\typeout{** loaded for the language `#1'. Using the pattern for}%
\typeout{** the default language instead.}%
\else
\language=\csname l@#1\endcsname
\fi
#2}}
\providecommand{\BIBdecl}{\relax}
\BIBdecl

\bibitem{Pea1988}
J.~Pearl, \emph{Probabilistic Reasoning in Intelligent Systems}.\hskip 1em plus
  0.5em minus 0.4em\relax Morgan Kaufmann, 1988.

\bibitem{MM2009}
M.~M{\'e}zard and A.~Montanari, \emph{Information, Physics, and
  Computation}.\hskip 1em plus 0.5em minus 0.4em\relax OUP, 2009.

\bibitem{OS2001}
M.~Opper and D.~Saad, \emph{Advanced Mean Field Methods: Theory and
  Practice}.\hskip 1em plus 0.5em minus 0.4em\relax {MIT} Press, 2001, nIPS
  workshop series.

\bibitem{sudderth2010nonparametric}
E.~B. Sudderth, A.~T. Ihler, M.~Isard, W.~T. Freeman, and A.~S. Willsky,
  ``Nonparametric belief propagation,'' \emph{Communications of the ACM},
  vol.~53, no.~10, p.~95, 2010.

\bibitem{DMM2009}
D.~L. Donoho, A.~Maleki, and A.~Montanari, ``Message-passing algorithms for
  compressed sensing,'' \emph{Proc. National Academy of Sciences of the United
  States of America}, vol. 106, no.~45, p. 18914, 2009.

\bibitem{Ran2011}
S.~Rangan, ``Generalized approximate message passing for estimation with random
  linear mixing,'' in \emph{Information Theory Proceedings, IEEE Internaional
  Symposium on}, 2011, p. 2168.

\bibitem{CR2005a}
E.~J. Cand\`{e}s and J.~Romberg, ``Signal recovery from random projections,''
  in \emph{Computational Imaging III}.\hskip 1em plus 0.5em minus 0.4em\relax
  San Jose, CA: Proc.~SPIE 5674, 2005, pp. 76--86.

\bibitem{KMS2012b}
F.~Krzakala, M.~M{\'e}zard, F.~Sausset, Y.~F. Sun, and L.~Zdeborov{\'a},
  ``Statistical-physics-based reconstruction in compressed sensing,''
  \emph{Physical Review X}, vol.~2, no.~2, p. 021005, 2012.

\bibitem{KMS2012}
------, ``Probabilistic reconstruction in compressed sensing: Algorithms, phase
  diagrams, and threshold achieving matrices,'' \emph{J. Stat. Mech.: Th. and
  Exp.}, no.~8, p. P08009, 2012.

\bibitem{donoho2012information}
D.~L. Donoho, A.~Javanmard, and A.~Montanari, ``Information-theoretically
  optimal compressed sensing via spatial coupling and approximate message
  passing,'' in \emph{Information Theory Proceedings (ISIT), 2012 IEEE
  International Symposium on}.\hskip 1em plus 0.5em minus 0.4em\relax IEEE,
  2012, p. 1231.

\bibitem{BayatiMontanari10}
M.~Bayati and A.~Montanari, ``The dynamics of message passing on dense graphs,
  with applications to compressed sensing,'' \emph{IEEE Transactions on
  Information Theory}, vol.~57, no.~2, p. 764, 2011.

\bibitem{CKZ2014}
F.~Caltagirone, F.~Krzakala, and L.~Zdeborov{\'a}, ``On convergence of
  approximate message passing,'' in \emph{Information Theory Proceedings
  (ISIT), 2014 IEEE International Symposium on}, 2014.

\bibitem{VS2013}
J.~P. Vila and P.~Schniter, ``Expectation-maximization gaussian-mixture
  approximate message passing,'' in \emph{Proc. 46th Annual Conference on
  Information Sciences and Systems}, 2012, p.~1.

\bibitem{RSF2014}
S.~Rangan, P.~Schniter, and A.~K. Fletcher, ``On the convergence of approximate
  message passing with arbitrary matrices,'' \emph{{arXiv preprint 1402.3210}},
  2014.

\bibitem{SAMP2014}
B.~\c{C}akmak, O.~Winther, and B.~H. Fleury, ``S-amp: Approximate message
  passing for general matrix ensembles,'' \emph{arXiv preprint 1405.2767},
  2014.

\bibitem{WJ2008}
M.~J. Wainwright and M.~I. Jordan, ``Graphical models, exponential families,
  and variational inference,'' \emph{Foundations and Trends in Machine
  Learning}, vol.~1, 2008.

\bibitem{TAP1977}
D.~J. Thouless, P.~W. Anderson, and R.~G. Palmer, ``Solution of `solvable model
  of a spin glass','' \emph{Philosophical Magazine}, vol.~35, no.~3, p. 593,
  1977.

\bibitem{BF2008}
E.~van~den Berg and M.~P. Friedlander, ``Probing the pareto frontier for basis
  pursuit solutions,'' \emph{SIAM Journal on Scientific Computing}, vol.~31,
  no.~2, pp. 890--912, 2008.

\bibitem{ZKM2013}
P.~Zhang, F.~Krzakala, M.~M{\'e}zard, and L.~Zdeborov{\'a}, ``Non-adaptive
  pooling strategies for detection of rare faulty items,'' in
  \emph{Communications Workshops, Proc. IEEE International Conference on},
  Budapest, Hungary, 2013, p. 1409.

\bibitem{BB2008}
P.~T. Boufounos and R.~G. Baraniuk, ``1-bit compressive sensing,'' in
  \emph{Proceedings of the $42^{\text{nd}}$ Annual Conference on Information
  Sciences and Systems}, Princeton, NJ, 2008, pp. 16--21.

\bibitem{JLB2011}
L.~Jacques, J.~N. Laska, P.~T. Boufounos, and R.~G. Baraniuk, ``Robust 1-bit
  compressive sensing via binary stable embeddings of sparse vectors,''
  \emph{arXiv preprint 1104.3160v3}, 2012.

\bibitem{KGR2012}
U.~S. Kamilov, V.~K. Goyal, and S.~Rangan, ``Message-passing de-quantization
  with applications to compressed sensing,'' \emph{IEEE Transactions on Image
  Processing}, vol.~60, no.~12, p. 6270, 2012.

\bibitem{XK2013}
Y.~Xu and Y.~Kabashima, ``Statistical mechanics approach to 1-bit compressed
  sensing,'' \emph{Journal of Statistical Mechanics: Theory and Experiment},
  no.~2, p. P02041, 2013.

\bibitem{XKZ2014}
Y.~Xu, Y.~Kabashima, and L.~Zdeborov{\'a}, ``Bayesian signal reconstruction for
  1-bit compressed sensing,'' \emph{arXiv preprint 1406.3782}, 2014.

\end{thebibliography}

\end{document}